\newcommand{\Tr}{\mathop{\rm Tr}} 
\def\Eq#1{Eq.~(\ref{#1})}
\def\pslash{\not \! P}
\def\Aslash{\not \!\! A}
\def\fwloop{\mathcal{W}}
\documentclass[12pt]{article}
\begin{document}
    \def\thepage{}
    %no page number on the title page
    
\begin{titlepage}
    
    % Definition of title page:
\title{Introducing Quarks in Confining Strings via the Fermionic Wilson Loop}
\author{\\
Vikram Vyas\footnote{Associate, Abdus Salam International 
Center for Theoretical Physics, Trieste, Italy.}
\footnote{E-mail: visquar@jp1.dot.net.in, visquare@satyam.net.in}\\
{ {\textit{Scientific Resource Center, }}} \\
{{\textit{The Ajit Foundation, Jaipur 302 018, India}}}
}
\date{}
\maketitle
\begin{abstract}
    In the world line representation of the fermionic effective
    action for QCD the interaction between Fermions and the 
    gauge field is contained in the fermionic Wilson loop, namely
    the Wilson loop for a spin-half particle. It is argued that a 
    string representation of the fermionic Wilson loop can
    provide a link connecting QCD with a dual description of a meson 
    as a quark and an antiquark connected by a string.
    This is illustrated by obtaining such a representation in 
    compact $U(1)$ gauge theories. The resulting description
    contains information about the interaction of the spin of the 
    quark with the world sheet degrees of freedom. Such interactions 
    may be of importance in the realization of chiral symmetry in the string 
    picture of QCD, and for delineating the possible presence of a 
    world sheet supersymmetry in QCD strings.
\end{abstract}

\end{titlepage}
\renewcommand{\thepage}{\arabic{page}}
\setcounter{page}{1}
\section{Introduction}\label{sec:1}

A very intuitive explanation of the observed confinement of quarks 
in hadrons is provided by the idea that hadrons are made of quarks 
which are connected by strings~\cite{tHooft1, Mandelstam}.  
Two approximations to QCD that suggest such a string picture of 
hadrons are the large $N$ expansion~\cite{tHooft2, WitteN}, 
where $N$ is the number of colors carried by the quarks, 
and the strong coupling expansion in the lattice 
regularization of QCD~\cite{Wilson74,KS75,PolyBook}. 
If we combine the insight from these two approximations then a very 
attractive and a plausible model of a meson as a quark and an antiquark 
connected by an elementary string is suggested. Such a description of 
QCD seems to imply that strings with quarks as endpoints are the dual 
variables to gluons and quarks. Assuming that such a duality does exist then the
non-perturbative phenomena of QCD, like confinement and chiral
symmetry breaking, should map into simple, 
calculable, perturbative phenomena of the dual theory. On the same
account, phenomena that have a simple perturbative description in QCD, like
deep inelastic scattering and the asymptotic freedom, could appear as 
non-perturbative phenomenon in the dual description. Thus, one can hope 
that QCD together with its dual description in terms of strings will 
allows us to obtain a qualitative and, an approximate, quantitative 
description of all the features of strong interactions.
Unfortunately, a quantitative transcription of this desired duality has remained 
elusive. The problems involved have been reviewed by 
Polchinski~\cite{Polchinski92}, and their possible resolution in view 
of the recent theoretical developments have 
been summarized by Polyakov~\cite{Polyakov97, Polyakov98, Polyakov00}. 

Most investigations of a string description of QCD do 
not include the quark degrees of freedom.
There are, of course, very good 
reasons for this neglect of fermionic degrees of freedom: We are far
from obtaining a string representation for pure  QCD and the addition of
fermionic degrees of freedom may make an already intractable problem
even more difficult. Also, one of the salient features of QCD which 
is captured by the strings, namely, confinement, can be studied
without introducing any fermionic degrees of freedom via the string
representation of Wilson loop~\cite{Wilson74}.

Yet, as has been emphasized in ref.~\cite{Polchinski92}, some of the 
most detailed evidence we have for the string picture of hadrons comes
from combining the ideas of dual resonance models with chiral 
symmetry~\cite{ademollo, lovelace} and one cannot talk meaningfully 
about chiral symmetry without introducing quark degrees of freedom. 
Thus, it is a possibility that the inclusion of fermionic degrees of freedom may
facilitate rather than hinder the task of obtaining a string description of hadrons.
More specifically, useful clues in developing a string representation 
for mesons 
might be obtained by considering the nature of pion and rho~\cite{Polchinski92}.
One would like to known, how does a string representation of mesons 
distinguish between a pion, which is a pseudo-scalar, approximate 
Nambu-Goldstone boson, and a rho which is a massive spin one 
particle~\cite{chiralstrings,lewellen}. 

The simplest implementation of the string picture of mesons would be to
consider models of strings whose end points carry spin and flavor 
quantum numbers of a quark. While it is known how to assign flavor 
quantum number to the end point of a string via Chan-Paton factors, 
there does not seem to be any simple  manner in which spin
degrees of freedom can be assigned to the end-points of an open
string~\cite{GreenBooks, StringBooks}. Alternatively, one could explore 
a model in which quarks are described as a point 
particle on which an open string terminates. This description 
essentially appears in the strong coupling expansion, whether one 
calculates the expectation value of a Wilson loop or when  one tries to 
diagonalize the lattice QCD Hamiltonian in the strong coupling
approximation~\cite{K1, K2}. The Wilson loop 
can be interpreted as the phase factor associated with the closed world 
line of a scalar test particle in the presence of an external gauge 
field~\cite{Wilson74, feynman3}. This suggests that one way of introducing 
the quark degrees of freedom in the string description of QCD 
is to look for a string 
representation of the fermionic Wilson loop (FWL), namely the phase factor
associated with a closed world 
line of a spin-half particle in the presence of an external gauge 
field.

One can construct a fermionic Wilson loop starting directly with the Lagrangian
of a spin-half particle~\cite{Berezin77, Brink76, Brink77, Barducci76}, an 
application of such an approach for studying studying chiral 
anomalies can be found in~\cite{LouisAG, Friedan}. This approach has 
many advantages, the symmetries are transparent and one can introduce 
flavor quantum numbers of the spin-half particle under consideration. 
The FWL also appears
naturally in the world line formalism of the fermionic Effective
action~\cite{Migdal, DHoker}, and it is this approach that will be 
used in the present preliminary investigation. The advantage of the latter 
approach is, as we will see in the next section, that it allows us to express, 
in the large N-limit,  all the mesonic Green's functions in terms of the expectation
value of the FWL. From the world line point of view
this approach gives us directly the appropriate vertex function for a 
pion or a rho.

Independent of our motivation of introducing quark degrees of freedom 
in the string description of meson, it has been suggested in 
ref.~\cite{Polyakov97} that even for obtaining the string representation of pure 
QCD the appropriate object is not the Wilson loop but the fermionic
Wilson loop\footnote{In refs.~\cite{Polyakov97, Migdal} the fermionic
Wilson loop is referred to as the SuperWilson loop, we have chosen to 
call it the fermionic Wilson loop in order to distinguish it from the
generalization of the Wilson loop to supersymmetric gauge 
theories~\cite{susyWL}.}.

Writing the mesonic propagators in 
terms of FWL leads us to the heart of the problem:
How to convert the expectation value of a FWL over the
gauge fields into a sum over surfaces whose boundary is the 
loop under consideration. In spite of much 
progress made in the case of supersymmetric Yang-Mill 
theories~\cite{Maldacena97,Maldacena98}, this problem remain unsolved for QCD. 
Still one can get some useful intuition by considering 
a string representation of theories where the mechanism of confinement is
well understood. In this paper we will use Polyakov's analysis of the 
string representation of Wilson loops in compact $U(1)$ gauge theory as
our starting point~\cite{Polyakov96}. The string 
representation of Wilson loops in compact $U(1)$ gauge theories can 
be be extended to the FWL, and as discussed in section
3, it does lead to a representation for ``mesons'' in which 
the string ends on the word line of quarks. Further, we will see in such a
representation there is an interaction between the spin degrees of 
freedom and the string world sheet mediated by a Kalb-Ramond field~\cite{KR}. 

In the last section of the paper we will discuss what can be abstracted 
from the present analysis of FWL in compact
$U(1)$ gauge theories to the case of interest which is QCD in the large N-limit. 
We will also outline the course of a future investigation of the question 
as to how the chiral symmetry is realized in a string representation 
and of the possibly related 
question of the existence of some form of world sheet 
supersymmetry for QCD strings~\cite{Polyakov97, horava}. 
%%%%%%%%%%%%%%%%%%%%%%%%%%%%%%%%%%%%%%%%%%%%%%%%%%%%%
\section{Mesons and the Fermionic Wilson Loop}\label{sec:2}
%%%%%%%%%%%%%%%%%%%%%%%%%%%%%%%%%%%%%%%%%%%%%%%%%%%%%
Consider the Euclidean partition function of an $SU(N)$ gauge theory
in the presence of an external source with the quantum numbers of a 
meson, say, a pion,  and with quarks in the fundamental representation, 
\begin{eqnarray}  
    Z[J] & = & \int D\psi D\bar{\psi} D A \exp \left \{ -
    \int_{x} (\frac{1}{4g^{2}} \Tr{F_{\mu \nu}F_{\mu \nu}} +
    \bar{\psi}(\pslash -\Aslash  - J 
     \gamma_{5}) \psi)\right \}
    \label{Z[j]}
\end{eqnarray}
$F_{\mu \nu}$ is the field strength tensor for the 
$SU(N)$ gauge theory, $\psi$ is the quark field, with flavor and 
color indices suppressed, $P = -i\partial$, $A $ is the matrix valued 
gauge field, $J$ is an external classical source, and  
the Feynman slash notation has been used. Fermionic degrees of freedom
can be formally integrated to give a functional integral over gauge 
fields,
\begin{eqnarray}
     Z[J] & = & \int D A \exp \left \{ -
    \int_{x} \frac{1}{4g^{2}} \Tr{F_{\mu \nu}F_{\mu \nu}} \right\} 
    \exp \left \{ -W[A, J] \right \},
    \label{eq:Z2}  \\
    -W[A, J] & = & \Tr \ln [\pslash -\Aslash  - J 
     \gamma_{5}].
    \label{eq:W}
\end{eqnarray}
Thus $Z[J]$ can be written as  a functional integral over gauge 
fields alone
\begin{eqnarray}
    Z[J] & = & Z[0] <\exp\{-W[A, J]\}>_{A},
    \label{eq:z}  \\
    <\exp\{-W[A,J]\} >_{A} & = & \frac{1}{Z[0]}\int D A \exp \left \{ -
    \int_{x} \frac{1}{4g^{2}} \Tr{F_{\mu \nu}F_{\mu \nu}} \right\} 
    \nonumber\\
    & &\times \exp\{-W[A,J]\}.
    \label{eq:avW}
\end{eqnarray}
We remind ourselves of an important simplification that takes place 
in the large $N$ limit~\cite{WittenCarges,makeenko}, namely
\begin{equation}
    <\exp\{-W[A,J] \}>_{A} = \exp\{-W[A^{m},J]\}
    \label{eq:masterfield}
\end{equation}
where $A^{m}$ is the so called ``master field''.
Thus one can write the pion propagator as
\begin{eqnarray}
    \Delta_{\pi}(x-y) & = & \left(\frac{\partial}{\partial J(x)} 
    \frac{\partial}{\partial J(y)} \right)_{J=0} \ln Z[J] \nonumber\\
    &=&\left(\frac{\partial}{\partial J(x)} 
    \frac{\partial}{\partial J(y)} \right)_{J=0} (-W[A^{m}, J])
    \label{eq:pionPropagator} 
\end{eqnarray}
This is of course the restatement of 
the familiar property of the large $N$ QCD~\cite{WitteN}, that the dominant 
diagrams are those with only a single quark loop which forms the 
boundary of the diagram.

The fermionic effective action, $W[A,J]$, has a natural representation in
terms of a world line path-integral of a Fermion\footnote{The world line 
formalism has been reviewed in~\cite{Schubert}, where additional 
references can  be found (see also~\cite{DHoker} and references 
there in).}. In what follows we will make use of the world line representation for 
$W[J,A]$ as developed by D'Hoker and Gagn\`{e}~\cite{DHoker}.
 First we note that one loop 
 diagrams involving even power of $\gamma_{5}$ are generated by the 
 real part of the fermionic effective action while the diagrams
 involving the odd powers of $\gamma_{5}$ are generated by the 
 imaginary part of the fermionic effective action~\cite{AGWitten} .
 Thus it is both natural 
 and convenient to consider separately the world line representation 
 of the real and the imaginary part of $W[A, J]$. The amplitude in 
 which we are interested, namely the pion propagator, involves two 
 insertions of $\gamma_{5}$, and thus is generated by the real part of 
 the fermionic effective action which has the following world line
 representation
 \begin{eqnarray}
     W_{R}[A, J] & = & \frac{1}{4}\int_{0}^{\infty} \frac{dT}{T} 
     \mathcal{N} \int_{PBC}DX \int_{APBC}D\psi  \Tr_{c} \mathcal{P}
     \nonumber \\
      &  & \times\exp \left\{ -\int_{0}^{T} d\tau \left( 
      \frac{\dot{x}^{2}}{2\mathcal{E}} + 
      \frac{1}{2}\psi_{A} \dot{\psi}_{A} + i  \mathcal{E}
      \psi_{\mu}\psi_{5}\partial_{\mu}J + \frac{1}{2} \mathcal{E}J^{2} \right) \right\}
     \nonumber \\
      &  & \times \fwloop[A;C],
     \label{RealW}
 \end{eqnarray}
where the fermionic Wilson loop, $\fwloop[A;C]$, is given by
\begin{equation}
    \fwloop[A;C] = \exp \left\{ i\int_{0}^{T} d\tau \left(\dot{x}\cdot 
    A - \frac{1}{2}\mathcal{E}\psi_{\mu}F_{\mu\nu}\psi_{\nu}\right)\right\},
    \label{FWL}
\end{equation}
with $\tau$ parametrizing the closed curve $C$, $\mathcal{E}$ is 
positive constant which can be though of as a constant einbein in 
the proper time gauge. In what follows we will work in the proper time 
gauge and scale the parameter $\tau$ so that $\mathcal{E}$ equals to 
identity. In \Eq{RealW} the trace over the gamma matrices has been replaced 
by path integral over anti-commuting variables, $\psi_{A}$, and 
$\Tr_{c}$ is the trace over color degrees of freedom. The index 
$A$ run over values one to six. When $A$ takes values from $1$ to $4$
then it is denoted by $\psi_{\mu}$. The world line anti-commutating 
variables $\psi_{A}$ are related to the six gamma matrices introduced 
by Mehta~\cite{Mayank}. For a detailed derivation of the
path-integral representation of the fermionic effective
action,~\Eq{RealW}, the reader is referred to ref.~\cite{DHoker}.

The interaction between spin-half particle and 
the gauge field $A_{\mu}$ is contained in the FWL,
\Eq{FWL}. The FWL is distinguished from the Wilson
loop,
\begin{equation}
    W[A;C] =\exp \left\{ i\oint_{c} d\tau \dot{x}\cdot A \right\},
    \label{WL}
\end{equation}
by the presence of the additional term 
$\psi_{\mu}F_{\mu\nu}\psi_{\nu}$ and is related to 
it via the area derivative~\cite{Migdal, makeenko},  
\begin{equation}
    \exp\left\{-\frac{i}{2}\oint d\tau 
    \psi_{\mu}(\tau)\psi_{\nu}(\tau)\frac{\delta}{\delta\sigma_{\mu\nu}} 
    \right\} W[C] = \fwloop[C].
    \label{WLtoFWL}
\end{equation}

From the above discussion we see that to obtain the desired string 
representation for mesons we need to express the expectation value 
of the FWL over the gauge fieldsas a sum over surfaces whose boundary 
is the loop $C$.  

It is interesting to note that 
FWL can be written in terms of a
superfield~\cite{Brink77,Friedan,Migdal}
\begin{equation}
    X_{\mu}(\tau, \theta) = x_{\mu}(\tau) + \theta\psi_{\mu}(\tau),
    \label{eq:superfield}
\end{equation}
where $\theta$ is the anti-commuting coordinate which together with 
$\tau$ parametrizes superspace. Using the covariant derivative on 
superspace,
\begin{equation}
    D = \frac{\partial}{\partial\theta} + 
    \theta\frac{\partial}{\partial\tau},
    \label{eq:superderivative}
\end{equation}
one can write FWL as
\begin{equation}
    \mathcal{W}[C] = \exp\left\{ \int d\tau d\theta DX_{\mu}(\tau, 
    \theta) A_{\mu}[X] \right\}
    \label{eq:superFWL}
\end{equation}
where the gauge field is function of the superfield,
\begin{eqnarray}
    A_{\mu}[X] & = & A_{\mu}\left(x_{\mu}(\tau) + \theta 
    \psi_{\mu}(\tau)\right)
    \nonumber  \\
     & = & A_{\mu}[x(\tau)] + \theta 
     \psi_{\nu}(\tau)\partial_{\nu}A_{\mu}[x(\tau)].
    \label{eq:superA}
\end{eqnarray}
The FWL is invariant under following local supersymmetric transformation
\begin{eqnarray}
    \delta x_{\mu} & = & \epsilon (\tau) \psi_{\mu},
    \nonumber \\
    \delta \psi_{\mu} & = & -\epsilon (\tau) \dot {x}_{\mu},
    \label{eq:susy}
\end{eqnarray}
where $\epsilon$ is an arbitrary infinitesimal Grassmann variable.
The reparametrization invariance and the supersymmetry of the
FWL have been studied in ref.~\cite{Migdal}.
These symmetries hint that corresponding string representation for the 
expectation value of the FWL may have a world sheet
supersymmetry, and the world sheet supersymmetry should reduce to 
the world line supersymmetry on the boundary. We will 
comment further on this possibility in the last section of the paper.
\section{String Representation of Fermionic Wilson Loop in Compact
$U(1)$ Gauge Theories}\label{sec:3}
In order to understand what additional information is contained in a 
string representation of the FWL, but not knowing the string 
representation for QCD, we look for a tractable mode as an alternative. 
Compact $U(1)$ in four dimensions, with an ultra-violet cutoff, is a confining 
theory~\cite{PolyBook, Polyakov77}. A string representation for this 
model was derived in ref.~\cite{Polyakov96}, and thanks to the simple 
relation between the Wilson loop and the FWL, \Eq{WLtoFWL}, we will 
be able to use this result to obtain a string representation for the FWL. 

We first summarize the results of ref.~\cite{Polyakov96}.
The expectation value of the Wilson loop,  $<W[A; C]>_{A}$, 
in a compact $U(1)$ gauge theory can 
be written as a sum over surfaces, 
$\Sigma_{c}$,  binding the curve $C$, 
\begin{equation}
<W[A\ ; C]>_{A} = \sum_{\Sigma_{c}}\exp\left\{ -S_{cs} [\Sigma_{c}]\right\} .   
\label{eq:wlstring}
\end{equation}
The confining string action, $S_{cs} [\Sigma_{c}]$, is given in terms 
of Kalb-Ramond field, $B_{\mu\nu}$, and can be written for 
large Wilson loops as
\begin{eqnarray}
    \exp\left\{ -S_{cs} [\Sigma_{c}]\right\} & = & \frac{1}{Z} \int D 
   B_{\mu\nu} \exp \left\{ -S_{KR[B]} + i\int_{\Sigma_{c}}
   d\sigma_{\mu\nu}B_{\mu\nu} \right\}, \label{eq:cstring} \\
    S_{KR}[B] & = & \int_{x} \left\{ \frac{1}{4e^{2}} 
   B^{2}_{\mu\nu} + f(dB)\right\},
    \label{eq:skr}  \\
     f(dB)  & = & \frac{1}{4\Lambda^{2}} \left ( \partial_{\mu} 
    B_{\nu\alpha}+\partial_{\nu}B_{\alpha\mu}+
    \partial_{\alpha}{B_{\mu\nu}}\right )^{2}.
    \label{eq:KRpot}
\end{eqnarray}
The antisymmetric tensor field $B_{\mu\nu}$ describes both the 
``photons'' and the magnetic monopoles which are present in compact 
$U(1)$ gauge theories, $\Lambda$ is an ultra-violet cutoff. The continuum 
limit of the confining string has been investigated in 
ref.~\cite{Trugenberg98,Trugenberg99}.

Since the field $B_{\mu\nu}$ appears quadratically in
\Eq{eq:cstring} it can be integrated out to obtain an non-local action 
for the confining string,
\begin{eqnarray}
    S_{cs} [\Sigma_{c}] = \int_{x, y} 
    \left\{T_{\mu\nu}(x) \Lambda^{2} G(x-y)T_{\mu\nu}(y) + 2 
    e^{2}\partial_{\nu} T_{\mu\nu}G(x-y)\partial_{\alpha}T_{\mu\alpha}\right\}
    \label{eq:cstringAction}
\end{eqnarray}
where $T_{\mu\nu}$ is defined in terms of the area element on the 
world sheet, $X_{\mu\nu}$, by
\begin{eqnarray}
     T_{\mu\nu}(x) & =  & \frac{1}{2}\int_{\sigma}X_{\mu\nu}(\sigma)\delta(x-x(\sigma))
    \label{eq:tmunu}  \\
    X_{\mu\nu} & = & \epsilon ^{ab}\frac{\partial 
    x_{\mu}}{\partial\sigma_{a}}\frac{\partial x_{\nu}}{\partial 
    \sigma_{b}}
    \label{eq:xmunu}
\end{eqnarray}
and $G(x-y)$ is the Yukawa Green's function~\cite{Trugenberg98}.

The non-local string action can be approximated by a local action by 
expanding Yukawa Green's function in inverse powers of the mass of
the Kalb-Ramond field~\cite{Polyakov96, Trugenberg98}
\begin{equation}
    S_{cs}[\Sigma] = c_{1} \Lambda^{2} \int_{\sigma} \sqrt{g} 
    -c_{2}\frac{\Lambda^{2}}{m^{2}} \int_{\sigma} \sqrt{g} 
    (\partial_{a}t_{\mu\nu})^{2} +  
    c_{3}me^{2}\oint_{C}d\tau
    \left(\frac{dx_{\mu}}{d\tau}\frac{dx_{\mu}}{d\tau}\right)^{\frac{1}{2}}
    \label{eq:confingstring}
\end{equation}
$c_{1}$, $c_{2}$  and $c_{3}$ are dimensionless constants depending on the regularization of 
the world sheet, $m$ is the mass of the Kalb-Ramond 
field, and $t_{\mu\nu} = \frac{1}{\sqrt{g}}X_{\mu\nu}$. 

Now we are in the position to obtain the string representation for 
the FWL. Applying the relation ship between the Wilson loop and the 
FWL as given by~\Eq{WLtoFWL} to the~\Eq{eq:cstring},
lead to:
\begin{eqnarray}
    <\fwloop[C]>_{A} & = & \frac{1}{Z_{KR}}\int DB \exp \left\{ -S_{KR}[B] 
    \right\}
    \nonumber  \\
     &  & \times \exp \left\{ i\int_{\Sigma_{C}} B_{\mu\nu}{d\sigma_{\mu\nu}
     -i\oint_{C}d\tau \psi_{\mu}B_{\mu\nu}\psi_{\nu}}
     \right\},
    \label{eq:fwlstringB}
\end{eqnarray}
where the action for the Kalb-Ramond field, $S_{KR}[B]$, is given 
by~\Eq{eq:cstring}. As an independent check, in the appendix we 
outline a derivation of the above representation, for the case of three 
dimensional compact $U(1)$ gauge theory, that does not use~\Eq{WLtoFWL}.

The string representation of the FWL is distinguished by
an additional boundary term involving the interaction 
between the spin degrees of freedom and the Kalb-Ramond field.
Again, as in the case of the Wilson loop, we can integrate out the 
Kalb-Ramond field to obtain a non-local string action for the 
FWL,
\begin{eqnarray}
    \mathcal{S}[\Sigma_{C}, C] & = & \int_{x,y}\left\{(T_{\mu\nu}(x) - 
    S_{\mu\nu}(x))D_{\mu\nu,\lambda\rho}(x-y)
    (T_{\lambda\rho}(y)-S_{\lambda\rho}(y)) \right\},
    \nonumber  \\
    S_{\mu\nu} (x) & = & 
    \oint_{C}d\tau \delta(x-y(\tau))\psi_{\mu}(\tau)\psi_{\nu}(\tau),
    \label{eq:smunu}  
\end{eqnarray}
where $D_{\mu\nu,\lambda\rho}(x-y)$ is the propagator for the 
Kalb-Ramond field. Expanding the above action,
and  substituting the expressions for $T_{\mu\nu}$ and $S_{\mu\nu}$ 
leads to
\begin{eqnarray}
    \mathcal{S}[\Sigma_{C},C]  =  \int_{\sigma_{1}, 
    \sigma_{2}}d\sigma_{\mu\nu}(\sigma_{1}) 
    D_{\mu\nu,\lambda\rho}(x(\sigma_{1})-y(\sigma_{2}))
    d\sigma_{\lambda\rho}(\sigma_{2})
    \nonumber  \\
      -2\oint_{\tau} d\tau \psi_{\mu}(\tau)\psi_{\nu}(\tau)\int_{\sigma} 
     D_{\mu\nu,\lambda\rho}(x(\tau)-y(\sigma))
     d\sigma_{\lambda\rho}(\sigma)
    \nonumber  \\
      + 
     \oint_{\tau_{1},\tau_{2}}d\tau_{1}d\tau_{2}\psi_{\mu}(\tau_{1})\psi_{\nu}(\tau_{1})
      D_{\mu\nu,\lambda\rho}(x(\tau_{1})-y(\tau_{2}))
      \psi_{\lambda}(\tau_{2})\psi_{\rho}(\tau_{2}).
    \label{eq:nlcfstring}
\end{eqnarray}
The first term in the above equation is the non-local form of the 
confining string action for the Wilson loop. The next two terms are 
specific to the FWL. The last term is purely a boundary term 
involving only Fermions. A similar term would appear even in a theory 
which has no strings, for e.g QED~\cite{worldlineQED}.

Of particular interest is the second term of ~\Eq{eq:nlcfstring}. It 
represents a non-local interaction between the Fermion at the 
boundary and the string world sheet. To make the nature of 
this interaction more transparent, we approximate the above non-local action by a 
local action using the derivative expansion for the 
propagator. Keeping only the leading terms,this leads to
\begin{eqnarray}
     \mathcal{S_{cs}}[\Sigma_{C},C] & = & S_{cs}[\Sigma_{c}] +
     c_{4}\Lambda^{2} \oint_{C}d\tau \psi_{\mu}(\tau) 
     \psi_{\nu}(\tau)t_{\mu\nu}(\tau) ,
\end{eqnarray}
where $S_{cs}[\Sigma_{c}] $ is the action for the confining string 
for the Wilson loop, \Eq{eq:confingstring}, and  $t_{\mu\nu}(\tau)$ is the value
of the area element density at the 
boundary of the surface.
%
%\footnote{We remind 
%ourselves that we are working in the proper time gauge for the 
%parameter $\tau$ which parametrizes the boundary, 
%and for general parameterization the last term in 
%above equation will take the form $\oint_{C}d\tau e(\tau)\psi_{\mu}(\tau) 
%\psi_{\nu}(\tau)t_{\mu\nu}(\tau)$ where $e(\tau)$ is the einbein.}. 
The leading effect of spin-string interaction is to correlate the 
spin of the ``quark'' with the area-element of the world sheet at the 
boundary. Since the world sheet in our model has no fermionic degrees 
of freedom, the spin-string interaction at the boundary breaks the 
world line supersymmetry of the FWL
\footnote{I would like to thank V.~P.~Nair for suggesting me to check 
this possibility.}. 
Our real interest of course is in the spin-string interaction for QCD. 
Even though we do not know the QCD string action, it is reasonable to 
expect that a term similar to the one we have delineated above should be 
present (though the nature and the number of world sheet degrees of 
freedom has to be different from those in the 
confining strings of the compact $U(1)$ gauge theories~\cite{Polchinski92}). 
Specifically, in the string representation of the FWL in QCD
the quark-string interaction should take place on the boundary. If 
the interaction between quark degrees of freedom and the world sheet 
variables is confined to the boundary, we are lead again to the 
question that was raised in the introduction, namely, how does the 
string communicate the relative spin orientation that distinguishes a 
pion from a rho. For a boundary interaction at the one end of an open 
string to influence the interaction at the other end would require a 
long distance correlation between the string variables on the world 
sheet. One would expect such a correlation to correspond to a 
massless mode in the interior theory. It is tempting to identify such a 
massless mode with a Nambu-Goldstone boson, this would be merely a 
tautology except that it would imply that the two ways of looking at the 
pion-rho mass difference, one as the consequence of a hyperfine 
splitting, and other as the result of the spontaneous breaking of chiral 
symmetry are related to each other in the string representation of QCD.
\section{Discussion}\label{sec:4}
Two points emerge from the previous sections that are independent of 
the model considered. Firstly, the FWL allows us to introduce the
required quark degrees of freedom in the string representation of 
QCD. Secondly, the string representation of mesons via the FWL
contains useful information about how the spin degrees of 
freedom of a quark and an antiquark interacts with the string that 
connects them. 

One of the motivation for studying the FWL is that its
string representation might shed some light on how chiral symmetry is 
realized in the dual picture of QCD. The FWL highlights
one issue regarding chiral symmetry breaking. 
If the right way of introducing quarks in the string 
representation of QCD is indeed via the FWL then the dynamics
responsible for chiral symmetry breaking must be located on the boundary 
of the string world sheet and not in the interior. As noted at the end 
of the last section such a mechanism also seems to suggest an intimate 
relationship between the spin-spin interaction, or the hyperfine 
splitting, in mesons and the existence of the Nambu-Goldstone bosons. 

The string representation of the FWL and the resulting 
spin-string interaction should be of even more importance if 
the (unknown) QCD string action has some form of world sheet 
supersymmetry. An evidence for the world sheet supersymmetry
emerges from the work of ref.~\cite{horava}, where it 
is shown that the string representation for QCD in two dimensions in the 
large $N$ limit, the t' Hooft model, has a rigid supersymmetry. Thus 
the t' Hooft model model offers an interesting laboratory for calculating and 
understanding the dynamics of the FWL. It is also a good place to 
test whether one can use the string representation of the FWL to 
calculate the spectrum of the theory. I hope to carry out these 
investigations in the near future.

\section*{Acknowledgments}
This work was started while visiting the
Physics Department, University of Utah, and the Physics Department, 
Brookhaven National Laboratory. I would like to thank Carleton~DeTar at 
the University of Utah, and Mike Creutz at the Brookhaven National 
Laboratory for making these visits possible. 
I would also like to thank the Associate Program of the 
Abdus~Salam International Center for Theoretical Physics (ICTP) 
which allowed me to visit ICTP. 
I have benefited from discussions with Madan Rao, Claudio Rebbi and 
Joseph Samuel, and would like to thank them for that. 
I am particularly grateful to Carleton~DeTar, Ansar~Fayyazuddin, 
V.~P.~Nair and Yong-Shi~Wu for their comments on a preliminary 
version of this paper.
%
%\newpage
\section*{Appendix}
The aim of this appendix is to outline the derivation of the string representation 
the fermionic Wilson loop in three dimensional compact $U(1)$ gauge theory.
The calculation is essentially a repetition of the calculation for the 
Wilson loop in ref.~\cite{Polyakov96} (see also ref.~\cite{details} for some 
of the details of the Wilson loop calculation.) The expectation value 
of the fermionic Wilson loop can be written as
\begin{equation}
    <\fwloop[C]>= <\fwloop>_{o}  <\fwloop>_{m}
    \label{eq:3du1}
\end{equation}
where the first factor is the gaussian integral,
\begin{equation}
    <\fwloop>_{o}= \frac{1}{Z[0]}\int DA 
    \exp\left\{-\frac{1}{4e^{2}}\int_{x}(\partial_{\mu}A_{\nu}-\partial_{\nu}A_{\mu})\right\}
     \fwloop,
    \label{eq:wzero}
\end{equation}
and the second factor, $<\fwloop>_{m}$, is the contribution from the 
magnetic monopoles. fermionic Wilson loop, $\fwloop$, is given
by~\Eq{FWL}. A monopole located at point $x$ gives following 
contribution to fermionic Wilson loop,
\begin{eqnarray}
    \fwloop^{1}_{m} & = & 
    \exp(-S_{m})\exp\left\{i(\frac{1}{2}\eta[x;C] -\chi[x;C])\right\},
    \label{eq:singlemonopole}  \\
    \eta[x;C] & = & \frac{\partial}{\partial 
    x_{\lambda}}\int_{\Sigma_{C}}d\sigma_{\lambda}\frac{1}{|x-y(\sigma)|},
    \label{eq:defeta}  \\
    \chi[x;C] & = & \frac{1}{4}\epsilon_{\mu\nu\lambda}\frac{\partial}{\partial 
    x_{\lambda}}\oint d\tau 
    \psi_{\mu}(\tau)\psi_{\nu}(\tau)\frac{1}{|x-y(\tau)|},
    \label{eq:defchi}
\end{eqnarray}
where $S_{m}$ is the action for a single magnetic monopole. Summing 
over all monopoles in the dilute gas approximation~\cite{PolyBook, 
Polyakov77} leads to
\begin{eqnarray}
     <\fwloop>_{m}& = & \frac{1}{Z_{m}}\int D\varphi 
    \exp\{-e^{2}\int_{x} [\frac{1}{2}(\partial \varphi)^{2}
    \nonumber  \\
     &  & + m^{2}(1-\cos(\varphi -\frac{1}{2}\eta[x;C] +\chi[x;C]))]\}
    \label{eq:fwlmonopole}
\end{eqnarray}
where $m^{2} = \exp {(-\frac{const}{e^{2}})}$.
Now consider the following functional integral
\begin{equation}
    W_{F}[C] = \frac{1}{Z} \int DB D\phi \exp\{-\Gamma[B, \phi; 
    \Sigma_{C}, C]\},
    \label{eq:stringansatz}
\end{equation}
where the effective action is given by
\begin{eqnarray}
    \Gamma & = & \int_{x} \left(\frac{1}{4e^{2}}B_{\mu\nu}^{2} - i 
    \epsilon_{\mu\nu\lambda}\partial_{\lambda}\phi B_{\mu\nu} + 
    e^{2}m^{2}(1-\cos\phi)\right)
    \nonumber  \\
     &  & + 2\pi i \int_{\Sigma_{C}} B_{\mu\nu}(\sigma) 
     d\sigma_{\mu\nu} -2\pi i \oint_{C} d\tau \psi_{\mu}(\tau) 
     \psi_{\nu}(\tau) B_{\mu\nu}(\tau).
    \label{eq:effaction}
\end{eqnarray}
Since the integration over field $B_{\mu\nu}$ is Gaussian we solve for 
the field $B_{\mu\nu}^{c}$ that minimizes the effective action 
$\Gamma$,
\begin{equation}
    B_{\mu\nu}^{c} = 
    ie^{2}\{\epsilon_{\mu\nu\lambda}\partial_{\lambda}\phi
    -2\pi\int_{\Sigma_{C}}\delta (x-y(\sigma))d\sigma_{\mu\nu}
    +2\pi\oint \psi_{\mu}(\tau) \psi_{\nu}(\tau) \delta 
    (x-y(\tau))d\tau
    \}.
    \label{eq:Bclass}
\end{equation}
Further using the following identity for the derivative of $\eta(x)$,
\begin{equation}
    4\pi\int_{\Sigma_{C}}\delta (x-y(\sigma))d\sigma_{\mu\nu} = 
    -\epsilon_{\mu\nu\lambda}\partial_{\lambda}\eta(x)
    -f_{\mu\nu}(x),
    \label{eq:etaderivatie}
\end{equation}
where,
\begin{eqnarray}
    f_{\mu\nu(x)} & = & \partial_{\mu}a_{\nu}-\partial_{\nu}a_{\mu}
    \label{eq:fmunu}  \\
    a_{\mu}(x) & = & \oint_{C}dy_{\mu}\frac{1}{|x-y|},
    \label{eq:amu}
\end{eqnarray}
and using a similar identity for the derivative of $\chi(x)$,
\begin{equation}
    2\pi\oint_{C}d\tau \delta (x-y(\tau)) 
    \psi_{\mu}(\tau)\psi_{\nu}(\tau) 
    =-\epsilon_{\mu\nu\lambda}\partial_{\lambda}\chi(x)
    -f_{\mu\nu}^{s}(x),
    \label{eq:chiderevative}
\end{equation}
where,
\begin{eqnarray}
    f_{\mu\nu}^{s}(x) & = & \partial_{\mu}b_{\nu} 
    -\partial_{\nu}b_{\mu}
    \label{eq:spinf}  \\
    b_{\mu}(x) & = & \frac{1}{2} 
    \oint_{C}\psi_{\mu}(\tau)\psi(\lambda)(\tau)
    \frac{\partial}{\partial 
    y_{\lambda}}\left(\frac{1}{|x-y(\tau)|}\right)
    \label{eq:aspin}
\end{eqnarray}
we get
\begin{equation}
    B_{\mu\nu}^{c}(x) = ie^{2}\left(\epsilon_{\mu\nu\lambda} 
    \partial_{\lambda}(\phi +\frac{\eta}{2}-\chi) 
    +(\partial_{\mu}A^{c}_{\nu}-\partial_{\nu}A^{c}_{\mu})
    \right)
    \label{eq:solB}
\end{equation}
where
\begin{equation}
    A_{\mu}^{c}(x) = \frac{1}{2}(\oint_{C}d\tau\left(
    \frac{dy_{\mu}}{d\tau}\frac{1}{|x-y(\tau)|} -
    \psi_{\mu}(\tau)\psi_{\lambda}(\tau)
    \frac{\partial}{\partial 
    y_{\lambda}}\left(\frac{1}{|x-y(\tau)|}\right) \right).
    \label{eq:Aclass}
\end{equation}
We notice that $A_{\mu}^{c}$ solves the integrand of the Gaussian 
integral, and that $B_{\mu\nu}^{c}$ has contribution both from 
magnetic monopoles and from the gauge field. Thus the functional integral over 
$B_{\mu\nu}$ reduces $W_{F}[C]$,~\Eq{eq:stringansatz}, to the 
product $<\fwloop>_{o}<\fwloop>_{m}$, once we identify the field 
$\varphi = \phi +\frac{\eta}{2}-\chi$ .
%%%%%%%%%%%%%%%%%%%%%%%%%%%%%%%%%%%%%%%%%%%%%%%

\end{document}